\documentclass[a4paper,11pt]{article}

\RequirePackage[OT1]{fontenc}
\usepackage{amsthm,amsmath,natbib,amssymb,enumitem,authblk}
\RequirePackage{hyperref}

\usepackage[dvipsnames]{xcolor}

\newcommand{\re}{\mathbb{R}}
\newcommand{\eg}{{\it e.g.}}
\newcommand{\ie}{{\it i.e.}}
\newcommand{\deffeq}{\mathrel{\overset{\makebox[0pt]{\mbox{\normalfont\tiny\sffamily def}}}{=}}}

\newcommand{\Feight} {\fontsize{8}{11}\selectfont  }

\newcommand{\Ften} {\fontsize{10}{11}\selectfont  }

\begin{document}
\pagestyle{myheadings}
\markboth{An introduction to Bent J\o{}rgensen's ideas}
                {G. M. Cordeiro, R. Labouriau and D. A. Botter}
\thispagestyle{empty}

\title{An introduction to Bent J\o{}rgensen's ideas}

\author[1] {\Feight Gauss M. Cordeiro}
\author[2] {Rodrigo Labouriau}
\author[3] {Denise A. Botter}
\affil[1]{Universidade Federal de Pernambuco, Brazil (gauss@de.ufpe.br)}
\affil[2]{Department of Mathematics, Aarhus University, Denmark (rodrigo.labouriau@math.au.dk)}
\affil[3]{Universidade de S\~ao Paulo, Instituto de Matem\'atica e Estat\'istica, Brazil (botter@ime.usp.br)}

\date{}

\maketitle

\begin{abstract}
We briefly expose some key aspects of the theory and use of dispersion models, for which Bent J\o{}rgensen played a crucial role as a driving force and an inspiration source. Starting with the general notion of dispersion models, built using minimalistic mathematical assumptions, we specialize in two classes of families of distributions with different statistical flavors: exponential dispersion and proper dispersion models. The construction of dispersion models involves the solution of integral equations that are, in general, untractable. These difficulties disappear when more mathematical structure is assumed: it reduces to the calculation of a moment generating function or of a Riemann-Stieltjes integral for the exponential dispersion and the proper dispersion models, respectively.
A new technique for constructing dispersion models based on characteristic functions is introduced turning the integral equations above into a tractable convolution equation and yielding examples of dispersion models that are neither proper dispersion nor exponential dispersion models. A corollary is that the cardinality of regular and non-regular dispersion models are both large.

Some selected applications are discussed including exponential families non-linear models (for which generalized linear models are particular cases) and several models for clustered and dependent data based on a latent L\'evy process.
\end{abstract}

\noindent
\Ften
{\bf Key-words:}
Dispersion models,
Exponential dispersion models,
Exponential family,
Non-linear models,
Proper dispersion models,
Saddlepoint approximations
\normalsize

\newpage

% ============================================================= %
\section{Introduction}
% ============================================================= %

The work of Bent J\o{}rgensen is rich and deep. Working on the foundations of generalized linear models (GLMs) he gave the main steps in the construction of the theory of exponential dispersion models  (EDMs), a class of parametric families of one-dimensional distributions. The necessity of expanding the EDMs to include further parametric families of distributions (\eg, the von Mises and the simplex distributions) led to the construction of the class of  proper dispersion  models (PDMs) in a process which culminated with the de\-ve\-lopment of the general theory of  dispersion models.  The class of dispersion models encompasses both the EDMs and the PDMs under the same umbrella. In this article, we will briefly expose some of the main aspects of the theoretical development referred above for which Bent J\o{}rgensen played a crucial role being a driving force and an inspiration source.

The path from the construction of EDMs to the development of the general theory of dispersion models presents an increasing level of abstraction. While EDMs rely on relatively strong assumptions (inherited from exponential families of distributions), the general dispersion models are based on a  minimalistic mathematical structure while keeping the desirable distributional and statistical properties. This theory-generation process was inductive in the sense that it represented a movement from several particular cases (disparate collections of distributions) to a general setup (large classes of families of probability measures with common properties). In this article, our exposition follows a reversed order, starting from the general theory of dispersion models (Section \ref{SS.DM}) involving a weak mathematical structure and specializing in two different statistical scenarios: EDMs and PDMs (Sections \ref {SS.EDM} and \ref{SS.PDM}, respectively).

We anticipate that when working in the general scenario, we meet some difficulties even for solving basic questions like finding a mathematical procedure to generate the families of probability involved there. For instance, the generation of dispersion models involves the solution of Fredholm integral equations that are known to be mathematically hard, and in many instances, it is difficult even to establish whether the equation has a solution or not. When introducing more mathematical structure, as expected, those difficulties gradually disappeared. Interestingly, arguments of probabilistic and statistical nature (\eg, a coincidence of an integral with the moment ge\-ne\-ra\-ting function, probabilistic properties of the characteristic function and exactness of the so-called $p^{\star}$ approximation) turn the hard mathematical problems tractable. The general integral equation becomes a convolution equation (solvable using standard methods of generalized functions or other methods of deconvolution)  or involves a simple calculation of a Riemman-Stieltjes integral in some of the more specific scenarios.

We stress that the deductive path used in our exposition (starting from the general and moving to more particular cases) was only feasible because we knew the entire inductive path referred to above and then we essentially reversed the order of the construction.  We believe that this choice of disposition of arguments is more comfortable to follow and enlighten the theoretical construction of the theory of dispersion models, but we invite the reader to imagine the reversed order while reading the text in order to get a feeling of the difficulties that she/he would face in constructing the general theory of EDMs. In our view, the most significant contribution that Bent J\o{}rgensen gave to the theory of dispersion models was to envision the structure that we briefly present here.

Bent J\o{}rgensen's professional trajectory was circular, starting in Denmark (Aarhus and Odense), he moved to several countries (England, Brazil and Canada) and eventually returned to Denmark (Aarhus-Foulum and finally Odense, where he was born). He obtained a master degree from Aarhus University (1979) with a thesis on the inverse Gaussian distribution,  under the supervision of Ole Barndorff-Nielsen. His master thesis was later published in a book \citep{Jorgensen1982}. Bent J\o{}rgensen received a Ph.D. title from the University of Southern Denmark (1987), also under the supervision of Ole Barndorff-Nielsen) and a Doctor of Science degree from Aalborg University (1997). He also studied at the Imperial College, in London, from 1981 to 1983. His famous book ``The Theory of Dispersion Models'' \citep{Jorgensen1997} was primarily derived from the results of his PhD thesis and his doctoral dissertation. A significant part of this work was published as lecture notes after teaching a course in the First School of Regression Models in S\~ao Paulo in 1989.

Bent J\o{}rgensen came to the ``Instituto de Matematica Pura e Aplicada'' (IMPA, in Rio de Janeiro, Brazil)  by the influence of Gauss Cordeiro who met him as a student at the Imperial College (London) in 1981. During a congress in York, in 1986, a few months before coming to Brazil, Bent expressed the desire to work in Brazil because of his Brazilian wife, Vera Botelho. Gauss Cordeiro took the steps to arrange that Bent came to IMPA since there he would have an environment suitable to do research without an overwhelming teaching and administrative load. He arrived in IMPA at the middle of the process of dissolution of the statistical group, but he fought to keep statistics alive there. From 1990 to 1992, Gauss Cordeiro stayed as a visiting professor at IMPA giving support to Bent. Together, they promoted an international event in asymptotic theory that attracted a large number of famous international statisticians.

Bent J\o{}rgensen visited the main statistical groups in Brazil during his stay from 1986 to 1992 in the country. In particular, he worked intensively with the group of statisticians at the ESALQ-USP through Clarice Dem\'etrio, with whom Bent kept a fruitful collaboration until his last days. This collaboration, including  some year latter  C{\'e}lestin C. Kokonendji and  John P. Hinde (among others), generated two main research lines: one on models for counts, see \cite{Bonat-etal2018}, and one on the relations between the so called Taylor's law 
(according to which the variances of responses related to some natural phenomena tend to be proportional to a power of the mean) and the Tweedie models, see  \cite{Jorgensen-etal2011}.

He recruited many students to work with him in IMPA (including Rodrigo Labouriau, Jos\'e Ra\'ul Martinez, among others). Bent attracted many international researchers in statistics as guests at IMPA including Ole Barn\-dorff-\-Nielsen, Preben Bl\ae{}sild, Michael S\o{}rensen, Jesper M\o{}ller (from Aarhus University), Ib Skovgaard (University of Copenhagen), Stephen Laurizen, S\o{}ren Lundbye-Christensen (Aalborg University) and Gerard Letac (from Paul Sabatier University). The departure of Bent J\o{}rgensen from IMPA to the University of British Columbia (UBC) in 1992 represented a significant loss for the Brazilian Statistics.

A group of statisticians was quickly gathered by Bent J\o{}rgensen at the UBC (\eg, S\o{}ren Lundbye-Christensen, as a recurrent visitor from Denmark,  and Peter Song, among others). They worked in a new research line in which EDMs were used to represent a latent stochastic process governing the temporal development of a phenomena of interest, see \cite{Jorgensen-etal1996A, Jorgensen-etal1996B, Jorgensen-etal1996C, JorgensenSong1997} and Section \ref{SS.latentProcesses}. In this period (around 1995--1997) Bent J\o{}rgensen worked also with Rinaldo Artes (currently at Insper - Instituto de Ensino e Pesquisa, S\~ao Paulo) developing part of his thesis entitled ``Extensions of generalized estimation equation theory to circular data and dispersion models'', which was approved in 1997 at University of S\~ao Paulo, see also  \cite{Artes-Jorgensen}.

Bent J\o{}rgensen left UBC and came back to Denmark in a temporary position shared between the Aarhus University (at the Department for Theoretical Statistics)  and the Biometry Research Unit at the Foulum Research Centre in 1996. In this period, he worked with Ole Barndorff-Nielsen (on simplex distributions and inferential separation techniques via concepts of sufficiency, ancillarity, and non-formation). The presence of Bent J\o{}rgensen brought much life to the statistical discussion in the Foulum group of statisticians led at that time by Rodrigo Labouriau.  He attracted Gordon Smyth who was working with dispersion parameter modeling and composite Poisson distributions, and latter  Antonieta Peres (whom met him in IMPA and worked with him on state space models, but she died a few years later), see \cite{ Botter-etal-2002}, and Renjun Ma, who was working with models of repeated measures, see \cite{Ma-Jorgensen2007,Ma-etal2009}. The last of Bent J\o{}rgensen's Ph.D. students was Wagner Hugo Bonat, who worked in multivariate extensions of dispersion models (currently at the Universidade Federal do Paran\'a) and Tweedie models, see \cite{BonatJorgensen2016} and  \cite{BonatKokonendjib}. All in all, Bent J\o{}rgensen worked in many different academic environments, spread over several countries, and always gathered many collaborators around himself. In the following sections, we review and discuss some of his main contributions. 

% ============================================================= %
\section{Dispersion Models}
% ============================================================= %

% ============================================================= %
\subsection{Introduction}  \label{SS.DM}
% ============================================================= %

The notion of dispersion models can be seen as a generalization of the normal distribution as we expose below. Consider the density of a univariate normal distribution with expectation $\mu\in\re$ and variance $\tau\in\re_+$,
\begin{equation}\nonumber
p(y; \mu, \tau) = (2\pi \tau)^{-1/2} \exp \left\{  -\frac{1}{2 \tau} (y-\mu)^2 \right\},
        \mbox{ for } y\in C =\re \, .
\end{equation}
Setting $d(y;\mu) =  (y-\mu)^2$ and $a(y;\tau) = (2\pi \tau)^{-1/2}$, we can express the density above in the form
\begin{equation}\label{eq.2.1.01}
p(y; \mu, \tau) = a(y;\tau) \exp \left\{ -\frac{1}{2 \tau}  d(y;\mu)  \right\},
       \mbox{ for } y\in C,
\end{equation}
where $d(y;\mu)$ is the squared Euclidean distance between the observation $y$ and the location parameter $\mu$.
The idea of dispersion models \citep[p. 4]{Jorgensen1997} is to replace the squared distance $d$ in (\ref{eq.2.1.01}) by another
sui\-table function, called the unit deviance, measuring how far an observation $y$ is from a central reference point $\mu$ of the distribution.
This idea turned to be rather fruitful since it generates a rich class of families of distributions, the dispersion models (to be defined precisely below), including many classic continuous, discrete and mixed type distributions. Dispersion models are often used in applications since they are families of distributions of outstanding statistical models as generalized linear models, exponential family nonlinear
mo\-dels (see Section \ref{SS.EFNLR}), generalized additive models, generalized linear mixed models, among others. Moreover, the dispersion models have some common mathematical properties that allow building an elegant and coherent theory of statistical inference and to construct models involving observations of well behaved stochastic processes (\eg, processes with stationary and independent increments, \ie, L\`{e}vy processes, see Section \ref{SS.latentProcesses}). In this article, we study the notion of univariate dispersion models and refer the interested reader to \cite{JorgensenLaurizen2000} and \cite{Jorgensen2013} for the multivariate case; see also \cite{BonatJorgensen2016}.

% ============================================================= %
\subsubsection{Basic Definitions}  \label{SS.DM.01}
% ============================================================= %

A dispersion model is a family of probability distributions parameterized by two parameters as follows. The starting point for the construction of the dispersion models is to define the concept of unit deviance. Let $S \subseteq \re$ be the set of the realizable values of the probability distributions contained in the family that we will construct (assumed to be the same for each element of the family). Denote the convex support of $S$ (\ie,  the smallest interval containing $S$) by $C$ and set $\Omega = int (C)$. Here $\Omega$ will be the parameter space of the referential parameter $\mu$. A \emph{unit deviance} is a function $d:C\times \Omega \rightarrow \re_+$ such that $d(y;y) = 0$ for all $y\in\Omega$, and $d(y;\mu ) > 0$ for all $(y,\mu) \in C\times \Omega$ such that $y\ne \mu$. A unit deviance $d$ is said to be  \emph{regular} when $d$ is continuously twice differentiable in $C\times \Omega$ and $\partial^2 d(\mu; \mu) / \partial \mu^2  >0$ for all $\mu$ in $\Omega$.

A {\it dispersion model} generated by a unit deviance $d$ is a parametric family of real probability measures with support contained in an interval $C\subseteq \re$ with density with respect to a suitable common dominating measure, $\upsilon$, taking the form (\ref{eq.2.1.01}). The dominating measure $\upsilon$ is typically the Lebesgue measure on $\re$ or the counting measure. Here the function $a:C\times \re_+ \rightarrow \re_+$ is such that the integral of the density $p$ is $1$. The parameters $\mu$ and $\tau$ are called the \emph{position parameter} and the \emph{dispersion parameter}, respectively. Classic examples of dispersion models are the normal ($d(y;\mu) = (y-\mu)^2$),
gamma ($d(y;\mu) = 2\{ y/\mu - \log(y/\mu) - 1\}$, for $y\in\re_+$),
von Mises  ($d(y;\mu) = 2\{ 1-\cos(y-\mu) \}$, for $y\in [0, 2\pi )$),
simplex ($d(y;\mu) =  (y-\mu)^2/\{ y(1-y)\mu^2(1-\mu)^2 \}$)
and Poisson ($d(y;\mu) =  2\{ y\log (y/\mu) - y + \mu \}$) distributions, see  \citep[p. 13-23]{Jorgensen1997} for further examples.

An immediate consequence of the above definition is that the position parameter $\mu$ is the mode of the density (\ref{eq.2.1.01}) when the dispersion parameter $\tau$ is fixed, since $\log \{ p(y; \mu, \tau)\}$ is proportional to $-d(y;\mu )$ and the function $d(y; \cdot )$ has by definition a minimum at $\mu = y$. A unit deviance can also be viewed as a generalization of the Kullback-Leibler information divergence   \citep{JorgensenLaurizen2000}.

We define for each regular unit deviance  $d:C\times\Omega \rightarrow \re_+$ the associated \emph{unit variance function}  $V:\Omega \rightarrow \re_+$ given by $V(\mu)= 2/\{ \partial^2 d(\mu;\mu)/ \partial\mu^2 \}$,  for each $\mu \in \Omega$. The unit variance function plays an important rule in the theory of dispersion models since it expresses the dependency of the variance on the expectation
under dispersion models and it characterizes uniquely the elements of some important classes of these models. Moreover, the variance function is useful to characterize certain forms of convergence of dispersion models, {\it c.f.}, \cite{Jorgensen1987A, Jorgensen1987B,Jorgensen1997}.

Two major classes of dispersion models will be studied in details: the proper dispersion models (PDMs) and the exponential dispersion models (EDMs). A dispersion model with density (\ref{eq.2.1.01}) is said to be a \emph{proper dispersion model} generated by a unit deviance $d:C\times\Omega \rightarrow \re_+$ when the function $a:C\times\re_+ \rightarrow \re_+$ factorizes as follows,
\begin{equation}\nonumber
a(y;\tau) = a_0(\tau)\,b(y),
 \mbox{ for all } y\in C \mbox{ and } \mu\in\Omega
\,  ,
\end{equation}
for a suitable choice of the functions $a_0$ and $b$. A PDM is said to be a \emph{regular proper dispersion model} when $C = \Omega$, the unit deviance $d$ is regular and $b(\cdot) = V^{-1/2} (\cdot)$, where $V$ is the unit variance function associated to the regular unit deviance $d$. There is essentially no loss of generality in assuming that $b(\cdot) = V^{-1/2}(\cdot)$ as we shall argue in Section \ref{SS.DM.03}.

A dispersion model generated by a  unit deviance  $d$ is said to be a \emph{reproductive exponential dispersion model} when the unit deviance has the form
\begin{equation}\label{eq.2.1.02}
d(y;\mu) = y f(\mu) + g(\mu) + h(y) \, , \mbox{ for all } y\in C \mbox{ and } \mu\in\Omega
\, ,
\end{equation}
for suitable functions $f,g$ and $h$. The special form of the unit deviance in (\ref{eq.2.1.02}) will be absorbed in the density 
(\ref{eq.2.1.01}) to obtain (in Section \ref{SS.EDM}, equation (\ref{exp})) an alternative representation of the probability density of the reproductive EDM which is the standard used in most of the literature on the subject.

Examples of reproductive EDMs are the normal, gamma, inverse Gaussian and Poisson distributions. The von Mises,  simplex, normal, gamma and inverse Gaussian distributions are classic examples of PDMs. Dispersion models that are neither a regular PDM nor a reproductive  EDM ``are still not well understood, mainly for lack of examples of this kind'' and because methods for ge\-ne\-ra\-ting those models are currently non-existent \citep[p. 8, last paragraph]{Jorgensen1997}. However, we give below an example and show a general method for obtaining this type of dispersion models. There are only three PDMs that are also reproductive EDMs: the normal, the gamma and the inverse Gaussian distributions \citep[Theorem 5.6, p. 188]{Jorgensen1997}; see also \cite{Daniels1980}.

% ============================================================= %
\subsubsection{Construction of Dispersion Models}  \label{SS.DM.02}
% ============================================================= %

We turn now to the problem of constructing dispersion models. Given a unit deviance $d$ defined on $C\times\Omega$, for a given real interval $C$ and $\Omega = int(C)$, a dispersion model can be obtained by finding a function $a:C\times\re_+\rightarrow\re_+$ such that the integral of the  right side of (\ref{eq.2.1.01}) is $1$, \ie, the function $a$ is the solution of the following Fredholm integral equation of first kind
\begin{equation}\label{eq.2.1.03}
\int_C a(y;\tau) \exp \left\{ -\frac{1}{2\tau}d(y;\mu) \right\}\upsilon (dy) = 1,
\mbox{for all } (\mu, \tau) \in \Omega\times\re_+ .
\end{equation}
The generation of a dispersion model involves then the construction of a real function from $C\times\Omega$ that is a unit deviance and finding
a solution of the related integral equation (\ref{eq.2.1.03}). Not all unit deviances generate a dispersion model since this equation
might not have a solution. Typically, solving this equation or even just establishing the existence and unicity for a solution is a hard problem. However, the integral equation (\ref{eq.2.1.03}) takes a simpler form for PDMs as we shall see in Section \ref{SS.PDMconstruction}.

Regarding the construction of a unit deviance, note that any distance on $C\times\Omega$ or any increasing function of a distance on $C\times\Omega$ is a unit deviance. The following construction based on basic properties of characteristic functions yields unit deviances with a tractable related integral equation. Let $P$ be a probability measure on $\re$ with characteristic function $\varphi$. If $P$ is symmetric around zero, then $\varphi (t) \in \re,$ for all $t\in \re$. Moreover, $\varphi (0) = 1$ and $\vert \varphi (t) \vert \le 1$ for all $t\in \re$\citep[p. 15]{Lucaks-1970}. Assuming further that $P$ is not a lattice distribution (\ie, $P$ is not concentrated on a set of the form $\{ a + nh, n = 0, \pm 1, \pm 2, \dots \}$ for some $a, h\in\re$ and $h>0$), then $\vert \varphi (t) \vert < 1$, for every $t\ne 0$ \citep[Theorem 1.1.3, p.2]{Ushakov1999}. Therefore, defining $d:C\times\Omega\rightarrow\re_+$ by $d(y;\mu)= 1 - \varphi (y- \mu ) = 1 - \varphi (\mu -y)$, for each $(y,\mu )\in C\times\Omega$ (the last equality follows from the symmetry of any characteristic function taking real values), yields a unit deviance. If we further require that the first two moments of $P$ are finite, then $d$ is twice differentiable and $\partial^2 / \partial \mu^2 d(\mu ; \mu )  =  1 + m_2 > 0$  (where $m_2$ is the second central moment of $P$), so $d$ is a regular unit deviance. For instance, the unit deviance given by  $d(y; \mu ) = 1 - \exp(-\vert y - \mu \vert)$, corresponding to the unit deviance constructed with the characteristic function of the Cauchy distribution, is not a regular unit deviance.

The integral equations (\ref{eq.2.1.01}) related to unit deviances constructed with characteristic functions as above, when $C=\re$, for a fixed $\tau = \tau_0$, becomes
\begin{eqnarray}\label{eq.2.1.06}
  1   =  \int_\re  a(y; \tau_0) \exp \left\{  -\frac{1 -  \varphi (\mu - y) }{2\tau_0}  \right\} \upsilon (dy)
 \! =    \!  \left [ a_{\tau_0} * K_{\tau_0} \right ] \! (\mu ),
\end{eqnarray}
for all $\mu\in\re$. Here the convolution operator, $``*"$,  refers to the convolution between functions. We want to solve the equation for $a_{\tau_0} (\,\cdot\,) = a (\,\cdot\,;\tau_0)$ where $K_{\tau_0}  (\,\cdot\, ) = \exp \{ -1/(2\tau_0)[ 1 - \varphi (\,\cdot\,) ]  \}$ is the kernel of a convolution equation. It is remarkable that the kernel $K_{\tau_0}$ is a characteristic function of a probability measure as proved in the  Corollary 1.3.4 in \cite{Ushakov1999}, page 8, and therefore it has a well defined Fourier transform. The solution of the convolution equation (\ref{eq.2.1.06}) does not depend on $\mu$, which is a consequence of the Lemma 5.2 in \cite{Jorgensen1997}.
A calculation involving the formalism of  tempered distributions, see \cite{Rudin-1973}, Chapters 7 and 9, and the delta Dirac distribution (in the sense of generalized functions), yields a general (formal weak) solution of the type
$$
a_{\tau_0} (\,\cdot\, )
 =
 \mathcal{F}^{-1} \left [ \delta  (\,\cdot\, ) /  \mathcal{F} \{   K_{\tau_0}  \} (\,\cdot\, )  \right ] 
  = \frac{1}{2\pi \int_\re \exp \left\{ -\frac{1}{2\tau_0}[1-\varphi(x)] \right\} dz } 
 .
$$
Here $ \mathcal{F}$ and $ \mathcal{F}^{-1}$  are the Fourier transform and the inverse of the Fourier transform, respectively. Note that the function $a(\,\cdot\, ; \tau) = a_{\tau} (\,\cdot\, )$ can be factorized as a product of a function of the observations $y$ and a function of the parameter $\tau$; therefore, the model generated is a PDM. 
A calculation involving projections on the orthogonal complement of $L^2(\re)$ closure of the $span\{ K(\,\cdot\, - \mu): \mu\in\re\}$ allows to construct, for each PDM generated as above, at least one DM that is neither a PDM nor an EDM.
% It is easy to see that the dispersion model generated by the unit deviance $d$ is not an EDM.
 Let $\mathcal{P}_0$ be the class of Borel probability measures in $\re$ that are symmetric around zero and are not a lattice distribution. The discussion above allows us to claim that the cardinality of the regular dispersion models that are not a PDM and not a reproductive EDM is at least the same as the cardinality of  $\mathcal{P}_0$. Moreover, the cardinality of those models that are regular is at least the cardinality of the set of elements of $\mathcal{P}_0$ that have the first two central moments finite.
To our best knowledge, this result has never been exposed in the literature before. This idea will be further explored in a future publication.

% ============================================================= %
\subsubsection{Some General Properties of Dispersion Models}  \label{SS.DM.03}
% ============================================================= %

Even though PDMs and reproductive EDMs have somewhat different distributional features, these two classes share some fundamental statistical properties, which are common to all regular dispersion models. We summarize these common properties below. First, we note that for any regular unit deviance $d$ it holds that
\begin{equation} \label{eq.2.1.07}
\frac{\partial^2d(\mu;\mu)}{\partial y^2} =
\frac{\partial^2d(\mu;\mu)}{\partial \mu^2}=
- \frac{\partial^2d(\mu;\mu)}{\partial \mu\partial y}, \mbox{ for all } \mu\in\Omega \, ,
\end{equation}
\cite[Lemma 1.1, p. 24]{Jorgensen1997}. This general result has two immediate consequences: it gives alternative ways to calculate the unit variance function and it implies that the unit deviance behaves similarly to the unit deviance of the normal family near its minimum, 
$\mu_0,$ since it follows from (\ref{eq.2.1.07}) that
\begin{equation} \nonumber
  d(\mu_0 + x\delta; \mu_0 + m\delta ) = \frac{\delta^2}{V(\mu_0)} (x-m)^2 + o(\delta^2) .
\end{equation}
This approximation sends us back to the initial idea  of viewing dispersion models as a form of generalization of the normal distribution.

A useful characteristic of dispersion models is a duality property for (well behaved) transformations, as we explain below.
Given a unit deviance $d$ on $C\times\Omega$ and a one-to-one function $f: C\rightarrow C_f$ ($C_f \subseteq \re$), the function
$d_f: C_f\times int(C_f) \rightarrow \re_+$
given  by $d_f= d \left (f^{-1}(z); f^{-1}(\xi) \right )$ for each $(z,\xi)\in C_f\times int(C_f)$
is also a unit deviance. Moreover, if a random variable $Y$ follows a dispersion model with unit dispersion $d$ and the function $f$ is monotone and differentiable, then the distribution of a transformed random variable $Z = f(Y)$ is in a dispersion model with density (in the continuous case)
\begin{equation}\nonumber
 p_Z(z; \xi, \tau ) =
  \frac{ a \left \{ f^{-1} (z), \tau  \right\}}
         { \vert f^{\prime}  \{ f^{-1} (z)  \} \vert }
         \exp \left\{ -\frac{1}{\tau}  d_f(z;\xi) \right\}
         \mbox{, for all } z\in C_f \, .
\end{equation}
In the discrete case, we eliminate the Jacobian  $1/ \vert f^{\prime}  \{ f^{-1} (z)  \} \vert$ of the above expression. The new dispersion model generated as above is called in the literature a \emph{re-parametrization by a transformation}, in an abuse of nomenclature since the new dispersion model is actually not a re-parametrization of the original dispersion model. If the unit deviance $d$ is regular and the transformation $f$ is twice continuously differentiable with $\vert f^\prime (y)\vert > 0$ for all $y\in\Omega$, then the unit deviance $d_f$ is also regular and has the associated unit variance $V_f(\xi ) = V\{f^{-1} (\xi ) \}[f^\prime \{f^{-1}(\xi)\}]^2$, where $V$ is the unit variance associated to the  unit deviance $d$. The transformation $f(y) = \int_{y_*}^y V^{-1/2} (v) dv$ (for a fixed $y_*$) has a constant variance function $V_*(\xi ) = 1$ for all $\xi\in int(C_f)$. This transformation, called the \emph{variance stabilizing transformation}, plays an important rule in the asymptotic theory of dispersion models.

The probability density of a regular dispersion model, $p$ given in (\ref{eq.2.1.01}), can be well approximated by
\begin{equation} \nonumber
 q(y;\mu, \tau) = \left\{ 2\pi\tau V(y) \right\}^{-1/2}
 \exp \left\{ -\frac{1}{2 \tau}  d(y;\mu)  \right\} \, ,
\end{equation}
in the sense that $p(y;\mu, \tau) /  q(y;\mu, \tau) \rightarrow 1$ when $\tau \rightarrow 0$ for each $y\in C$ and $\mu \in\Omega$.
This approximation is called the \emph{saddlepoint approximation}. Clearly, this convergence is equivalent to
$a(y,\tau)/  \{ 2\pi\tau V(y) \}^{-1/2} \rightarrow 1$ when $\tau \rightarrow 0$. The saddlepoint approximation is often very accurate and it is useful because the function $a(\cdot,\cdot)$ in (\ref{eq.2.1.01}) is difficult to be calculated or numerically evaluated.
Note that the integral of the functions $q(\cdot;\mu,\tau)$ are not necessarily $1$, therefore we define the \emph{renormalized saddlepoint approximation} by $q_0(\cdot;\mu,\tau )= q(\cdot;\mu,\tau)\,a_0(\mu,\tau)$, where $a_0(\mu,\tau) = 1/ \int_C q (y ;\mu,\tau ) \nu(dy)$. The approximation $p(y;\mu, \tau)  \sim   q_0(y;\mu, \tau)$ (as $\tau \rightarrow 0$) is often more accurate than the original saddlepoint approximation.
 
% ============================================================= %
\subsection{Exponential Dispersion Models} \label{SS.EDM}
% ============================================================= %

An EDM is defined as a class of real distributions having a density with respect to a suitable dominating measure taking the form
\begin{equation}
p(y;\theta,\tau)=\exp\left\{\frac{1}{\tau}\Big[y\theta
-b(\theta)\Big]+c(y;\tau)\right\}. \label{exp}
\end{equation}
Here $b:\Theta \rightarrow \re$, and $c: C \times \re_+  \rightarrow \re$ are known appropriate functions, $\Theta$ is an open set in $\re$, $\theta\in \Theta$ and $\tau>0$ are called the {\it canonical parameter} and the {\it dispersion parameter}, respectively. Typically, the dominating measure is the Lebesgue measure in $\re$, yielding continuous distributions, or the counting measure, generating discrete distributions. A {\it natural exponential family} is obtained when the dispersion parameter $\tau$ is kept fixed. The notion of EDMs was pioneered by \cite{Tweedie1}, who studied several special cases and pointed out important structural properties. The theory of EDMs was systematically exposed in \cite{Jorgensen1987A, Jorgensen1997} where several mathematical properties of EDMs were presented for the first time.
The terminology ``exponential dispersion model" reflects the exponential form of the density of those distributions and the important role played by the dispersion parameter $\tau$.

The cumulant generating function (cgf) of a distribution with density given by (\ref{exp}) is
\begin{equation}\label{cgf}
K(t;\theta,\tau)=\frac{1}{\tau}\Big[b(\theta+ \tau t)-b(\theta)\Big],
\end{equation}
which depends only on the function $b$, termed the {\it cumulant generator}. 
Setting $b(\theta)=\theta^2/2$, $b(\theta)=-\log(-\theta)$ and $b(\theta) = -(-2\theta)^{1/2}$
yield the normal, gamma and inverse Gaussian distributions, respectively. The discussion above implies that we can construct one EDM for any specified non-degenerate function $b$ since this function uniquely determines the class of cgf of an EDM via (\ref{cgf}) and the cgf uniquely determines the distribution (an instance of the Fourier inversion theorem for the characteristic functions). Therefore, there exist many EDMs.

Differentiating (\ref{cgf}) yields the expectation of the distribution with density  (\ref{exp}) given by $\mu \deffeq M(\theta) \deffeq \partial b(\theta)/\partial \theta$. The function $M$, associating each value of $\theta$ to the expectation of the corresponding distribution, is called the {\it mean value mapping}. By inverting the mean value mapping (we show below that $M$ is indeed invertible) we obtain that $\theta= M^{-1} (\mu ) =  b^{\prime-1}(\mu)\deffeq q(\mu)$. The variance of the distributions given by  (\ref{exp}) are then
$\tau V(\mu)$, where the {\it variance function} $V(\,\cdot\,)$ has the following alternative forms, for each value of $\mu$,
$$
V(\mu)=
\frac{\partial^2b(\theta)}{\partial\theta^2}=
\frac{\partial b^{\prime}\{q(\mu)\}}{\partial\theta}=
\frac{\partial M\{ q(\mu)\}}{\partial\theta}=
m^{\prime}\{q(\mu)\}>0.
$$
Consequently, the mean value mapping $\mu$ is a strictly increasing function and the parameter $\theta=q(\mu)=\int V(\mu)^{-1}d\mu$ is a known one-to-one function of $\mu$. For a given variance function $V(\,\cdot\,)$, we can easily obtain the inverse of the mean value mapping $q(\,\cdot\,)$ and then calculate $b(\theta)=\int q^{-1}(\theta) d\theta$ (for each $\theta$) and the cgf  given by (\ref{cgf}). The elements of the EDMs are then uniquely determined by their variance functions $V(\,\cdot\,)$. Moreover, the variance functions play a key role for studying several of the structural properties of EDMs.

The EDMs with quadratic variance function given by $V(\mu)=a \mu^2+b \mu+c$ form an important and well studied class of EDMs.  \cite{Morris} studied those EDMs in details and proved that this subclass is composed of only six distributions: the normal ($a=b=0, c=1$), the gamma
($a=1, b=c=0$), the Poisson ($a=c=0, b=1$), the binomial ($a=-1, b=1, c=0$), the negative binomial ($a=b=1, c=0$) and the generalized secant hyperbolic ($a=c=1, b=0$) distributions. J\o{}rgensen (1997) discussed further the generalized secant hyperbolic by taking $b(\theta)=-\log\{\cos(\theta)\}$ and
$$c(y;\tau) = \log \left[\frac{2^{(1-2\tau)/\tau}} {\tau \Gamma(\tau^{-1})} \right]
 - \sum_{j=1}^{\infty} \log \left[1 + \frac {y^2} {(1+ 2 j \tau)^2}\right],$$
where $\Gamma(\cdot)$ is the gamma function.

The $r$th \normalsize cumulant of a distribution contained in an EDM with density given by (\ref{exp}) is $\kappa_r=\kappa_r(\theta,\tau)=\partial^{r}K(t;\theta,\tau)/\partial^r t \big{|}_{t = 0}$. Therefore, the function $b(\theta)$ generates all cumulants of the distribution
$\kappa_r=\tau^{r-1} \partial^{r}b(\theta) / \partial\theta^r$ for $r \ge1$. A direct implication of that is that any distribution contained in an EDM has finite cumulants of all orders, which rules out the use of EDM for modeling situations where it is required that the distribution has very heavy tails.

\cite{Tweedie1} proved the normal convergence of the random variable $Z=(Y-\mu)/\sqrt{\tau}$, where the distribution of the random variable $Y$ is contained in an EDM,   using an expansion for its cgf. The cgf of $Z$ follows from (\ref{cgf}) as
$$K_Z(t;\theta,\tau)=-\mu \frac{t}{\sqrt{\tau}}+\frac{1}{\tau}\,\bigg[b(\theta+ \sqrt{\tau} t)-b(\theta)\bigg].$$
By expanding $b(\theta+ \sqrt{\tau} t)$ in a Taylor series around zero with the cumulants of $Y$ as coefficients and collecting like terms,
we obtain

$$K_Z(t;\theta,\tau)=V(\mu)\, t^2/2 +\sum_{k=3}^{\infty}\,\frac{\partial^{k}b(\theta)}{\partial^{k}\theta}\,\frac{\tau^{k/2-1}\,t^k}{k!}.$$
Based on this expansion, we conclude that
\begin{equation}\label{asy}
Z=(Y-\mu)/\sqrt{\tau}\,\,\stackrel{\rm D}{\rightarrow}\,\,\text{N}(0,V(\mu))\,\,\,\,\,\text{when}\,\,\,\,\,\tau\rightarrow 0,
\end{equation}
where $\stackrel{\rm D}{\rightarrow}$ denotes convergence in distribution. Clearly, equation (\ref{asy}) is exact for the normal distribution since the derivatives of $b(\theta)$ of order greater than two vanish.

Equation (\ref{asy}) generalizes a number of known results on convergence to normality such as those for the gamma and inverse Gaussian
distributions. In fact, under some regularity conditions, distributions contained in  EDMs  are approximate normally distributed for small values of $\tau$. In this way, the standard asymptotic theory applies for small values of the dispersion parameter as well as for large sample sizes. The mathematical conditions for this result called {\it small dispersion asymptotics} were fully discussed in J\o{}rgensen (1987b).

Let $Y_1,\cdots,Y_n$ be independent and identically distributed (iid) random variables with distribution contained in an EDM 
with mean $\mu$ and dispersion parameter $\tau$. The form of the cgf in (\ref{cgf}) implies that the distribution of the sample
mean $\bar{Y}=\sum_{i=1}^n Y_i/n$ belongs to the same EDM as the distribution of the elements of the sample with  mean $\mu$ and dispersion parameter  $\tau/n$. This result includes well-known convolution properties of the normal, gamma and inverse Gaussian distributions and implies that the distributions contained in EDMs are infinite divisible provided the dispersion parameter is allowed to take values arbitrarily close to zero.

% ============================================================= %
\subsubsection{Tweedie Models}
% ============================================================= %

A important sub-class of the models defined by (\ref{exp}), called {\it Tweedie models} is obtained when $V(\mu)=\mu^{p}$, for $p\in (-\infty, 0]\cup [1,+\infty)$. A full discussion of these distributions with varying $p$ was first addressesed by \cite{Tweedie1,Tweedie2} and further discussed by \cite{Jorgensen1987A}. The support of the Tweedie models depends on the value of $p$. The normal, Poisson, gamma and inverse Gaussian distributions can be obtained from (\ref{exp}) for $V(\mu)=\mu^{p}$ and $p=0,1,2$ and $3$, respectively. Distributions generated by extreme stable distributions are determined when $p<0$ with support on $\mathbb{R}$. J\o{}rgensen (1987a) demonstrated that there exist no EDM with power va\-riance functions  for $0 < p <1$. When $1 < p < 2$, we obtain the Gamma compound Poisson distributions which are inte\-res\-ting because they are continuous for $y > 0$ but have positive probability mass at zero. For $p > 2$ ($p \ne 3$), we obtain continuous distributions generated by positive stable distributions. When $p$ increases to $+\infty$ the Tweedie models converge to extreme stable distributions.

We define the cumulant generator $b_{p}(\theta)$ for Tweedie models ($p\neq1,2$) by
$$b_{p}(\theta) = (2-p)^{-1} \left[(1-p)\,\theta \right]^{\frac {p-2} {p-1}}.$$
Further, $b_{1}(\theta)=\exp(\theta)$ and $b_{2}(\theta)=-\log(-\theta)$.

The unit deviance $d_p(\cdot ;\cdot)$  of a Tweedie model follows by the following straightfoward calculation
$$
d_{p}(y;\mu) = 2\,\int_{\mu}^y \frac{(y-t)}{V(t)} dt =
2 \left\{\frac{[max(y,0)]^{2-p}}{(1-p)(2-p)}-\frac{y \mu^{1-p}}{1-p}+\frac{\mu^{2-p}}{2-p}\right\}.
$$
The remaining quantities in equations (\ref{exp}) can be evaluated numerically for the cases $1<p<2$ and $p>2$ in power series following
\cite{Jorgensen1987A}. See, also \cite{BonatKokonendjib}.

% ============================================================= %
\subsubsection{Saddlepoint Approximations}
% ============================================================= %

We adopt the notation $K_Y^{(j)}(t;\theta,\tau)=\partial^j K_Y(t;\theta,\tau)/\partial t^j$ for $j \ge 1$.
The saddlepoint approximation for the density of $Y$ takes the form
\begin{equation}\label{saddle}
\pi_Y(y;\theta,\tau) \simeq \left[\frac{1}{2\pi\,K_Y^{(2)}(\hat{\lambda};\theta,\tau)}\right]^{1/2}\,\exp\left[
K_Y(\hat{\lambda};\theta,\tau)-\hat{\lambda} y\right],
\end{equation}
where the saddlepoint $K_Y(\lambda;\theta,\tau)-y$ is found by solving the (usual nonlinear) equation $K_Y^{\prime}(\hat{\lambda};\theta,\tau)=y$.

This equation was derived by Daniels (1954) as an approximation for any density given its cgf, although it can be applied here to approximate
the density of an EDM when $\tau$ is small. In fact, the $r$th cumulant of $Y$ is of order $O(\tau^{r-1})$ (for $r \ge1$) and then the
density of $Y$ with large precision parameter $1/\tau$ can be considered as the density of the sample average in large samples.
By differentiating (\ref{cgf}), we find $b^{\prime}(\theta+\tau\hat{\lambda})=y$ and then $\hat{\lambda}
=[q(y)-\theta]/\tau$. By definition of the variance function, we have $K_Y^{(2)}(\hat{\lambda};\theta,\tau)=\tau\,V(y)$. Inserting these quantities in the last density approximation, the saddlepoint approximation for the ED density can be expressed in a simple form
\begin{equation}
\pi_Y(y;\theta,\tau)\simeq \left[\frac{1}{2\pi\,\tau\,V(y)}\right]^{1/2}\exp\left[-\frac{1}{2 \tau}\,d(y;\mu)\right],\label{spoint}
\end{equation}
which holds when $\tau\rightarrow 0$. This approximation is exact independent of $\tau$ for the normal distribution since the third and fourth cumulants of $Y$ vanish. Equation (\ref{spoint}) is equivalent to the asymptotic result from (\ref{eq.2.1.01}): $\sqrt{\tau}\,a(y;\tau) \rightarrow [2\pi\,V(y)]^{1/2}$ when $\tau \rightarrow 0$.

The distribution function of $Y$ follows approximately from (\ref{spoint}) as
\begin{equation*}
\Pi_Y(y;\theta,\tau)=P(Y \le y)\simeq \int_{-\infty}^y \left[\frac{1}{2\pi\,\tau\,V(x)}\right]^{1/2}\,\exp\left[-\frac{1}{2 \tau}\,d(x;\mu)\right]\,dx.
\end{equation*}

By integrating the last equation, Lugannani and Rice (1980) proved that
\begin{equation*}
\Pi_Y(y;\theta,\tau)\simeq \left[\Phi\left(\frac{r}{\sqrt\tau}\right)+\sqrt\tau\,\phi\left(\frac{r}{\sqrt\tau}\right)\,\left(\frac{1}{r}-\frac{1}{u}\right)\right],
\end{equation*}
where $\Phi(\cdot)$ and $\phi(\cdot)$ are the standard normal distribution and density, respectively, $r=\rm{sgn}(y-\mu)\,\sqrt{d(y;\mu)}$ is called the {\it deviance residual} and $u=\frac{V(y)^{1/2}}{2}\,\frac{\partial d(y;\mu)}{\partial y}$ is the {\it dual score residual}. This equation is very easy to be applied to compute probabilities for any EDM since it involves only the variance and the deviance functions.

We now move to the density of the sample average $\overline{Y}=\sum_{i=1}^{n}Y_i/n$ of iid random variables $Y_1,\cdots,Y_n$ having density (\ref{exp}) and cgf (\ref{cgf}). The density function of $\overline{Y}$ follows from the Fourier inversion integral as
$$\pi_{\overline{Y}}(y;\theta,\tau)=\frac{1}{2 \pi}\, \exp\Big[- \rm{i} t\, y + n \,  K_Y( \rm{i} t/n;\theta,\tau)\Big],$$
where $\rm{i}=\sqrt{-1}$. This equation is suitable for Daniels' saddlepoint approxi\-mation. Setting $z =\rm{i} t/n$, the saddlepoint of 
$K_Y(z;\theta,\tau)- z y$ is $K^{\prime}_Y(\hat z;\theta,\tau)=y$. Then, the density approximation of $\overline{Y}$ can be expressed as

\begin{eqnarray*}
\pi_{\overline{Y}}(y;\theta,\tau)&=&\Bigg[\frac{n}{2\pi\,K_Y^{(2)}(\hat z;\theta,\tau)}\Bigg]^{1/2}\,\exp\Big\{n \,\Big[K_Y(\hat z;\theta,\tau)
-\hat{z} \,y\Big]\Big\},
\end{eqnarray*}
which can provide good results in practice.

It is much more frequent in statistical applications to compute distribution functions than density functions. By integrating the last equation, the cumulative distribution function (cdf) of $\overline{Y}$ has the form
\begin{equation*}
\Pi_{\overline{Y}}(y;\theta,\tau)\simeq \large{\int_{-\infty}^y}\Bigg[\frac{n}{2\pi\,K_Y^{(2)}(t;\theta,\tau)}\Bigg]^{1/2}\,
\exp\Big\{n\,\Big[K_Y(t;\theta,\tau)- t \,x\Big]\Big\} dx,
\end{equation*}
where $t= t(x)$ is determined by $K_Y^{\prime}(t;\theta,\tau)=x$. By transformation of variables and integration with respect to the saddlepoint variable $t$ instead of $x$, we obtain $K_Y^{(2)}(t;\theta,\tau)\,dt=d x$ and then
\begin{equation*}
\Pi_{\overline{Y}}(y;\theta,\tau)\simeq \large{\int_{-\infty}^{t(y)}}\Bigg[\frac{n K_Y^{(2)}(t;\theta,\tau)}{2\pi}\Bigg]^{1/2}\exp\Big\{n \Big[K_Y(t;\theta,\tau)-t \,K_Y{\prime}(t;\theta,\tau)\Big]\Big\} dt,
\end{equation*}
where $t(y)$ is found by solving $K_Y^{\prime}(t(y);\theta,\tau)= y$. This integral for $\Pi_{\overline{Y}}(y;\theta,\tau)$ is much easier
to compute than the previous one because it includes explicitly the saddlepoint function in the integrand. The saddlepoint approximation for the cdf of ${\overline{Y}}$ follows from Lugannani and Rice (1980) as
\begin{equation}\label{sadlemean}
\Pi_{\overline{Y}}(y;\theta,\tau)\simeq \Phi\left[r(y)\right]+\phi\left[r(y)\right]\,\left[\frac{1}{r(y)}-\frac{1}{u(y)}\right],
\end{equation}
where
$$r(y)=\rm{sgn}[t(y)]\,\left\{2 n\, \left[y\,t(y)-K_Y(t(y);\theta,\tau)\right]\right\}^{1/2}$$
and
$$u(y)=t(y)\,\left[n\,K_Y^{(2)}(t(y);\theta,\tau)\right]^{1/2}.$$
Equation (\ref{sadlemean}) provides highly accurate results for the probabilities associated
with $\overline{Y}$.

% ============================================================= %
\subsection{Proper Dispersion Models} \label{SS.PDM}
% ============================================================= %

We consider now the notions of general dispersion models and general PDMs that will allow us to understand some statistical inferential aspects of the notion of PDMs introduced in Section \ref{SS.DM}. A {\it general dispersion model} is a family of real distributions parameterized by two parameters,  $\theta\in \Theta \subseteq \re$ and $\lambda \in \Lambda \subseteq \re_+ ,$ where $\Theta$, $\Lambda$ are intervals, $\Lambda$ is unbounded to the right and the density (or Radon-Nikodym derivative) with respect to a common dominating real measure $v$ is of the form
\begin{equation}\label{eq.2.3.1.01}
p(y; \theta, \lambda) = a(y;\lambda) \exp \left\{\lambda t(y;\theta) \right\},
       \mbox{ for all } y\in S,\ \theta\in \Theta\ \mbox{and}\ \lambda\in \Lambda .
\end{equation}
Here $S$ is the support of the dominating measure $v$, and $a: S\times\Lambda \rightarrow \re_+ ,
$ and $t: S\times\Theta \rightarrow \re_+$ are suitable functions. When the function $a$ factorizes as $a(y;\lambda)=a_0(\lambda)b(y)$, for all $y\in S$ and $\lambda\in \Lambda$, the family of distributions above is said to be a {\it general proper dispersion model} (general PDM). Note that when the function $-d$ is a unit dispersion, $\lambda = 1/\tau$ and $\mu = \theta$, then the general dispersion model and the general PDM coincide with the dispersion models and the PDMs defined in Section \ref{SS.DM}. Moreover, (\ref{eq.2.3.1.01}) coincides with (\ref{eq.2.1.01}) defined with the unit deviance $t$ instead of $d$. Apparently, the setup above is much more general than the situation considered in Section \ref{SS.DM} since here  we are not requiring the function $t$ to be a unit deviance. However, we will argue that in order to obtain families of distributions with some desirable statistical inferential properties we will need to introduce some restrictions in the general definition given above that will render the present definition essentially equivalent to the setup discussed in Section \ref{SS.DM}. In this way, we are using an embedding of classes of families of distributions to obtain a better understanding of inferential properties of the statistical models based on PDMs.

% ============================================================= %
\subsubsection{Some Key Properties}
% ============================================================= %

The notion of yoke and yokable function defined below will allow us to connect the notions of PDMs (as defined in  Section \ref{SS.DM}) and general dispersion models (defined above). Moreover, these notions will allow us to characterize the existence of maximum likelihood estimates (MLEs) for PDMs. Given an interval $\Omega \subseteq \re ,$ a function $t:\Omega\times\Omega \rightarrow \re$ is said to be a {\it yoke} if $\sup_{\begin{subarray}{l}\theta\in\Omega\end{subarray}} t(y;\theta) = t(y;y)$ for all $y\in \Omega$. When additionally, $t(y;y)=0,$ for all $y\in\Omega$, then the function $t$ is called a {\it normed yoke}.
If $d:\Omega\times\Omega \rightarrow \re_+$ is a unit deviance, then $-d$ is a normed yoke.
On the other hand, if the function $t:\Omega\times\Omega \rightarrow \re$ is a yoke and we denote for each $y\in\Omega$ the supremum $\sup_{\begin{subarray}{l}\theta\in\Omega\end{subarray}} t(y;\theta)$ by
$\hat t(y)$, then the function given by $\tilde t (y;\theta) = \hat t(y )  - t(y;\theta)$ is a normed yoke.

A function $t:C\times C \rightarrow \re$, where $C$ is a real interval, is said to be \emph{yokable} when the following three conditions are satisfied:
{\it i)}  $\sup_{\begin{subarray}{l}\theta\in\Omega\end{subarray}} t(y;\theta) < \infty$  for all $y\in C$;
{\it ii)} There exists an open interval $\Omega\subseteq C$ such that, for each $y\in\Omega$ the supremum $\hat\theta_y = \sup_{\begin{subarray}{l}\theta\in\Omega\end{subarray}} t(y;\theta)$ is unique; and
{\it iii)} The function $\hat\theta: \Omega \rightarrow int(\Theta)$ given by $\hat\theta (y) = \hat\theta_y$ (for each $y\in\Omega$) is a bijection.

If the function $t$ used in (\ref{eq.2.3.1.01}) to define general PDMs is yokable, then we might define a PDM using the unit deviance $d:C\times\Omega \rightarrow \re$ given by
$d(y;\mu) = 2 \left [\hat t(y) - t \left\{ y; \hat\theta (\mu ) \right\} \right]$, for all $y\in C$ and $\mu\in\Omega$.
In this case, the density of a general PDM takes the form
\begin{equation}\label{eq.2.3.1.02}
p(y; \theta, \lambda) =
a_0(\lambda)b(y)
 \exp \left\{\lambda\hat\, t(y) - \frac{\lambda}{2}
d(y; \mu)\right \},\\
 \end{equation}
for all $y\in S$, $\mu\in \Theta$ and $\lambda\in \Lambda$.
Here $a(y;\lambda)  = a_0 (\lambda) \exp\{\lambda \hat{t}(y)\}$.
If we further assume  that  $\Omega = S = C$, that the function $d$ is regular unit deviance, and that for all $y\in \Omega$, $b(y)=V^{-1/2}(y)$, then the family of distributions defined by  (\ref{eq.2.3.1.02})  is  a regular PDM as defined in Section \ref{SS.DM.01}.
We argue below that it is natural from the statistical point of view to assume the function $t$ in  (\ref{eq.2.3.1.01}) to be yokable.
First, the conditions {\it i)} and {\it ii)} of the definition of yokable function ensure the existence of at least a local maximum of the likelihood function for $\mu$, obtained when fixing the index parameter $\lambda$. Note that this is a minimal necessary requirement for the existence and uniqueness of the maximum likelihood estimates. The condition  {\it iii)} implies that $\theta$ is a re-parametrization of the position parameter of the related proper dispersion model.

In order to elucidate some basic statistical properties of general dispersion models, let us consider the so called  Barndorff-Nielsen's $p^*$ formula  \citep{Barndorff-Nielsen-1983, Barndorff-Nielsen-1988} for approximating the conditional distribution of the MLE (for a given statistical model) given an ancillary statistic (\ie, a statistic carrying no information on the parameter of interest, see  \cite{JorgensenLabouriau2012}, Chapter 2 for details). Here we consider the maximum likelihood estimation of the parameter $\mu$ in a statistical model defined by  (\ref{eq.2.3.1.02}) when the parameter $\lambda$ is kept fixed and the estimation is based on a single observation, $y\in S$. Taking a degenerate ancillary statistics (\eg , a constant statistic) the $p^*$ formula yields an approximation to the marginal distribution of the MLE. Since in these circumstances the MLE of $\mu$ is  $\hat{\mu}(y)=y$ and the $p^*$ approximation is defined by
\begin{equation}\label{eq.2.3.1.03}
p(y; \theta(\mu), \lambda) \sim p_0(y;\mu ,\lambda),
\end{equation}
where $p_0$ is the renormalized saddlepoint approximation corresponding to the unit deviance $d$, with
$p_0(y; \mu, \lambda) =  a_0(\mu,\lambda) V^{-1/2 } (y)
 \exp \left\{- \frac{\lambda}{2} d(y; \mu)\right\}$,
for all $ y\in S$, $\mu\in \Theta$ and $\lambda\in \Lambda$.
The Barndorff-Nielsen's formula is said to be {\it exact} if the two sides of (\ref{eq.2.3.1.03}) coincide for all $y$ in $S$ and all $(\mu, \lambda )$ in $\Omega\times\Lambda$.

A consequence of the saddlepoint approximation above is that standard dispersion models  are asymptotically normal distributed for $\lambda$ large. In this way, the Barndorff-Nielsen's formula may be viewed as a refinement of the normal approximation to the distribution of the MLE. Furthermore, there is a strong result for regular PDMs stating that (assuming $b$ continuous at $y=\mu$) the following three statements are equivalent: i) The Barndorff-Nielsen's formula is exact for all $\lambda\in \Lambda$; ii) the Barndorff-Nielsen's formula is asymptotically exact in the sense that the ratio $p_0/p$ tends to 1 as $\lambda \to \infty$ for all $y$ and $\mu$ in $\Omega$; iii) The function $\hat{t}(y)$ is constant on $\Omega$, and $b(y) \propto V^{-1/2}(y)$ (see \cite{Jorgensen1997}, Theorem 5.4 and Corollary 5.5). When these statements hold, the normalizing constant $a_0(\mu,\lambda)$ does not depend on $\mu$, and satisfies $a_0(\mu,\lambda)\propto a(\lambda) \exp\left\{\lambda \hat{t}(\mu)\right\},$ and $a(\lambda)\sim \sqrt{\frac{\lambda}{2\pi}}\exp\left\{-\lambda \hat{t}(\mu)\right\}$ as $\lambda \rightarrow \infty .$
In conclusion, the density in (\ref{eq.2.3.1.02}) defines a regular PDM when the Barndorff-Nielsen's formula is exact.

We discuss below some other properties of PDMs which will show some of the peculiarities of those families of distributions.
For any fixed value of the position parameter $\mu$, say $\mu = \mu_0$, the family given by (\ref{eq.2.1.01}) is an exponential family
with canonical statistic $d(\cdot;\mu_0)$ and canonical parameter $\tau$. This is enough to recognize that the general form of the density of a distribution belonging to the exponential family is obtained when setting the position parameter $\mu$ fixed.

Another property of PDMs is that when the dispersion parameter is fixed, say  $\tau = \tau_0,$ the unit deviance is a pivotal statistic for 
$\mu$.  That is, if  $Y$ is a random variable having density (\ref{eq.2.3.1.02}) with respect to $\upsilon$, then the distribution of the random variable  $d(Y,\mu)$ does not depend on $\mu$. This property follows by observing that the integral
$\int_C b(y) \exp \left\{  - d(y,\mu) / (2 \tau_0) \right\}  \upsilon (dy )= 1/a_0(\tau_0)$ does not depend on the value of $\mu$; it is easy then to prove that the moment generating functions of the random variables $T = T_\mu= d(Y,\mu)$, for $\mu\in\Omega$, are all equal and depend only on $\tau_0.$ This key result is the Lemma 5.2 in  \cite{Jorgensen1997}.

% ============================================================= %
\subsubsection{Construction of proper dispersion models}  \label{SS.PDMconstruction}
% ============================================================= %

As discussed in Section \ref{SS.DM.02}, given a unit deviance an associated dispersion model can in principle be constructed by solving the integral equation (\ref{eq.2.1.03}), but this is in general a hard problem.  However, that integral equation takes the following simpler form for PDMs
 \begin{equation}\label{eq.3.2.01}
  a_0(\tau)\int_C b(y) \exp \left\{ -\frac{1}{2\tau} d(y;\mu) \right\} \upsilon (dy) =
  1,
 \mbox{ for all } (\mu, \tau) \in \Omega\times\re_+
 \, ,
\end{equation}
which has a solution $a_0(\tau) =  1 /  \int_C b(y) \exp \left\{ -d(y;\mu) /2\tau \right\} \upsilon (dy)$, provided that the integral involved exists. The integral $\int_C b(y) \exp \left\{ -d(y;\mu) /2\tau \right\} \upsilon (dy)$  does not vanish (since the integrand is positive apart from a $\upsilon$-null set) and does not depend on the parameter $\mu$ because of the second key property of PDMs discussed before.

Another useful technique for constructing PDMs involves the use of a transformation group, say $G$, acting freely and transitively
on $\Omega = C = \re$; see Section 3.3 of \cite{JorgensenLabouriau2012} for the basic definitions and a short account on transformation groups in statistical inference. Here we denote the action of $G$ by $(g,y) \mapsto gy$ for $(g,y) \in G\times \Omega$. Let $t,b:\Omega\rightarrow\re_+$ where the function $b$ is invariant by the action of $G$(\ie, $b(gy) = b(y)$ for all $g\in G$ and $y\in\Omega$) and $t$ is an arbitrary function. Assume, moreover that $\int_\Omega b(y) \exp \left\{ \lambda t( g^{-1} y) \right\} dy < \infty$, for $\lambda$ in an  interval 
$\Lambda \subseteq \re_+$  that is unbounded from the right. Since $G$ acts freely and transitively on $\Omega$, there is a one-to-one correspondence between $\Omega$ and $G$. Assume further that the supremum $\hat t = \sup_{g\in G} t(g^{-1} y)$ is finite. Then, it is easy to see that $t(g^{-1} y)$ is yokable and we might define the unit deviance $d(y;\mu) = 2 \left\{ \hat t - t(\hat g^{-1}_\mu y )  \right\}$, where $g^{-1}_\mu$ is the MLE of $g$ when the observation is $\mu$. The type of dispersion model constructed with this kind of unit deviance is called a \emph{transformation dispersion model}.

 Examples of the construction techniques based on transformation groups described above are the special PDMs called \emph{location-dispersion model} for which the density function with respect to the do\-mi\-nating measure $\upsilon$ has the form
 $ p(y; \mu, \tau) = c(\tau) \exp \left\{ - d(y-\mu)/(2\tau) \right\}$,
 where $C = \Omega = \re$ and $d$ is a unit deviance;  here the transformation group is the group of translations with action 
$y \mapsto (g + y)$ . Another example is the von Mises distribution defined using the transformation group of rotations of the unit circle with action $y \mapsto module \{ 2\pi; (g + y)\}$.

% ============================================================= %
\section{Applying Dispersion and Exponential Dispersion Regression}
% ============================================================= %

We review the Exponential Dispersion (ED) regression to extend the well-known GLMs, discuss some improved hypotheses tests and
some models for clustered and dependent data based on latent L\'evy processes.

% ============================================================= %
\subsection{Exponential Family Non-Linear Models}
\label{SS.EFNLR}
% ============================================================= %

We consider models where the random variables $Y_1,\cdots,Y_n$ are assumed independent and each $Y_i$ has a density or probability function of the form (\ref{eq.2.1.01}) with mean $\mu_i=E(Y_i)$ on a convenient support.

We define the ED regression by the random component (\ref{eq.2.1.01}) and the systematic component
\begin{equation}\label{systematic}
g(\mu_i)=\eta_i=f(\boldsymbol{x}_i;\boldsymbol{\beta}),
\end{equation}
where $g(\cdot)$ is a known one-to-one twice continuously differentiable link function, $\eta_i$ and 
$\boldsymbol{x}_i = (x_{i1}, \cdots, x_{ip})^T$
denote the linear predictor and the $p\times 1$ vector of non-stochastic independent variables associa\-ted with the $i$th observation, respectively, $\boldsymbol{\beta}=(\beta_1,\cdots,\beta_p)^T$ is a $p$-vector of unknown parameters, and $f(\cdot;\cdot)$
is a (possibly nonlinear) twice continuously differentiable function with respect to $\boldsymbol{\beta}$. Here a standard GLM
is obtained when the function $f(\cdot;\cdot)$ is bi-linear, i.e., when $f(\boldsymbol{x}_i;\boldsymbol{\beta}) = \boldsymbol{x}_i^T \boldsymbol{\beta}$.

The systematic component relates the explanatory variables $\boldsymbol{x}_i$ to the mean parameter $\mu_i$ of
interest. The $n\times p$ matrix of derivatives of $\boldsymbol{\eta}$ with respect to $\boldsymbol{\beta}$, specified by 
$\widetilde{\bf{X}}=\widetilde{\bf{X}}(\boldsymbol{\beta})=\partial\boldsymbol{\eta}/\partial\boldsymbol{\beta}$,
is assumed to have rank $p$ for all $\boldsymbol{\beta}$. We have $p+1$ parameters to be estimated: the vector $\boldsymbol \beta$ and $\tau$. The ED regression model has two important components: the ED class for the response variable and a possible nonlinear regression on a vector 
$\boldsymbol{\beta}$ by means of the link function. We assume that the standard regularity conditions for likelihood theory hold. The ED regression model was called {\it exponential fa\-mi\-ly nonlinear} (EFNL) model by Cordeiro and Paula (1989), thus extending the well-known idea of the GLMs by allowing a nonlinear regression structure for the explanatory variables. Wei (1998) wrote an excellent book for the EFNL models.

Let $\boldsymbol{y}=(y_1,\cdots,y_n)^T$ be a vector of observations and $\ell=\ell(\boldsymbol{\beta},\phi)$ be the total
log-likelihood function for a given ED regression model expressed in terms of $\boldsymbol{\beta}$ and $\tau$. A simple calculation shows
that $E(\partial^2\ell/\partial\tau\partial\boldsymbol{\beta})=0$, i.e., the parameters $\boldsymbol{\beta}$ and $\tau$ are globally orthogonal.
Let $\widehat{\boldsymbol{\beta}}$ and $\hat{\tau}$ be the MLEs of $\boldsymbol{\beta}$ and $\tau$, respectively. Let $\mu_i = g^{-1}(\eta_i)$ be the inverse link function evaluated at the linear predictor. Given a data vector $\boldsymbol{y}$, the total deviance for the ED regression is defined as $$D(\boldsymbol{y};\boldsymbol{\mu})= \sum_{i=1}^n d(y_i;\mu_i).$$

The vector $\widehat{\boldsymbol{\beta}}$ can be calculated by minimizing the total deviance $D(\boldsymbol{y};\boldsymbol{\mu})$
with respect to this parameter vector. The MLE of $\boldsymbol{\beta}$ does not depend on the dispersion parameter $\tau$.
Let $\widehat{\boldsymbol{\beta}},\widehat{\boldsymbol{\eta}}$ and $\widehat{\boldsymbol{\mu}}=g^{-1}(\widehat{\boldsymbol{\eta}})
=\left(g^{-1}(\hat \eta_1),\cdots, g^{-1}(\hat \eta_n) \right)^T$ be the MLEs of the vector of regression coefficients $\boldsymbol{\beta}$, the vector of linear predictors $\boldsymbol{\eta} = (\eta_1, \cdots, \eta_n)^T$ and the vector of means $\boldsymbol{\mu} =(\mu_1, \cdots, \mu_n)^T$, respectively. The Fisher information matrix for $\boldsymbol{\beta}$ is $K(\boldsymbol{\beta})=\tau\,\widetilde{\bf{X}}^{T} \textrm{W}\widetilde{\bf{X}}$, where $\textrm{W}={\rm diag}\{w_{1},\cdots, w_{n}\}$ is a diagonal matrix with weights $w_{i}=V(\mu_{i})^{-1}(\partial\mu_{i}/\partial\eta_{i})^{2}$.

The algorithm to estimate $\boldsymbol{\beta}$ can be carried out by using iteratively re-weighted least squares (IRLLS)
$$\widehat{\boldsymbol{\beta}}=(\widehat{\widetilde{\bf{X}}}^{T}\widehat{\textrm{W}}\widehat{\widetilde{\bf{X}}})^{-1}
\widehat{\widetilde{\bf{X}}}^{T}\widehat{\textrm{W}}\widehat{\boldsymbol{z}},$$
where $\widehat{\widetilde{\bf{X}}}$ and $\widehat{\textrm{W}}$ are the quantities $\widetilde{\bf{X}}$ and $\textrm{W}$
evaluated at $\widehat{\boldsymbol{\beta}}$, $\widehat{\boldsymbol{z}}=(\hat z_{1},\cdots, \hat z_{n})^{T}$ is the working vector
with components $z_{i}=\eta_{i}+(y_{i}-\mu_{i}) \partial\eta_{i}/\partial\mu_{i}$ at $\widehat{\boldsymbol{\beta}}$. These nonlinear equations have the same form as the estimating equations for GLMs with a local model matrix $\widetilde{\bf{X}}$ instead of a known design matrix and can be solved by iterative methods. The IRLLS algorithm is easily implemented using some standard statistical software such as SAS or the GAMLSS
script in {\bf R} (R Development Core Team, 2007).

Some asymptotic results for the ED regression were obtained by Cordeiro and Paula (1989), Cordeiro and McCullagh (1991) and Simas and Cordeiro (2009), among others, which produce wider results than those for GLMs.

Estimation of the parameter $\tau$ is a more difficult problem than the estimation of $\boldsymbol{\beta}$ and the complexity depends entirely on the functional form of $c(y;\tau)$. The MLE $\widehat\tau$ is a function of the deviance of the model, namely $\widehat\tau$  is the solution of the following equation
\begin{equation}
\widehat{\tau}^{2}\left. \sum_{i=1}^{n}\frac{\partial c(y_{i};\tau)}
{\partial\tau }\right|_{\tau=\widehat{\tau}}= \left[\sum_{i=1}^n l(y_i;y_i) -\frac{D(\boldsymbol{y};\boldsymbol{\mu})}{2}\right].\label{estimatephi}
\end{equation}

Equation ($\ref{estimatephi}$) requires in general the use of a nonlinear algorithm to compute numerically $\hat\tau$ except for normal and inverse Gaussian models. However, for some ED regressions, the form of $c(y;\tau)$ is complicated, and $\hat\tau$ could be difficult to compute from (\ref{estimatephi}). In these cases, we can use a moment estimate of $\tau$ directly from $\widehat{\boldsymbol{\mu}}$ given by
$\widetilde\tau = \frac{1}{n-p} \sum_{i=1}^{n} (y_i-\hat\mu_i)^{2}V(\hat\mu_i)^{-1}$ on the grounds that the expected value of $D(\boldsymbol{y};\boldsymbol{\mu})/\tau$ is approximately $n-p$ for a well fitted model.

If ($\ref{exp}$) is a two-parameter full exponential family with canonical para\-meters $1/\tau$ and $\theta/\tau$, the following
decomposition holds
\begin{equation}
c(y;\tau)=\frac{1}{\tau} a(y)+d(\tau)+e(y)\label{decompos}
\end{equation}
and then explicit expressions for $\widehat\tau$ are possible. Clearly, equation (\ref{decompos}) is valid for normal, gamma and inverse Gaussian distributions but does not hold for all ED distributions in (\ref{exp}).

The above results apply to all GLMs by setting $f(\widetilde{\bf{X}};\boldsymbol{\beta})=\bf{X} \boldsymbol{\beta}$ for a given design matrix 
$\bf{X}$. Several diagnostic measures for the ED regression are simple extensions of those measures for GLMs.

The Tweedie regression models are extensively used in several areas for non-negative right-skewed data and continuous data that allow zero observations. Bonat and Kokonendjib (2017) proposed maximum likelihood, quasi-likelihood and pseudo-likelihood methods for estimation and inference of Tweedie regression models with unknown power parameter $p$ in the va\-rian\-ce function. The last two methods are fast and computationally simple because they employ the first two moments only and thus it does not require the function $c_p(y;\tau)$.

% ============================================================= %
\subsection{Improved Tests  }
\label{SS.IT}
% ============================================================= %

Bartlett and Bartlett-type corrections improve the large-sample $\chi^2$ approximation to the null distribution of the likelihood ratio, score and gradient statistics, when the sample size is finite. For a detailed discussion, see \cite{Cordeiro-Cribari}. In several papers, like those from  \cite{Ferrari-Cordeiro-Cribari}, \cite{CordeiroPaulaBotter} and \cite{Medeiros-Ferrari-Lemonte}, for example, improved likelihood ratio, score and gradient tests, respectively, were obtained in the class of dispersion models.

% ============================================================= %
\subsection{Modelling Clustered and Dependent Data}
% ============================================================= %

% ============================================================= %
\subsubsection{Latent Stochastic Processes Based Models}
\label{SS.latentProcesses}
% ============================================================= %
Dispersion and ED mo\-dels were used to study data containing clustered and dependent observations
\citep{Jorgensen-etal1996A,
          Jorgensen-etal1996B,
          Jorgensen-etal1996C,
					JorgensenSong1997,
          JorgensenTsao1999,
					Jorgensen-etal1999,
          Artes-Jorgensen,
          Botter-etal-2002,
          Ma-Jorgensen2007,
          Ma-etal2009} in recent years.
The common idea explored there is that the dependence in the data is modeled using a latent stochastic process, the observations being conditionally independent given the latent process. For example, in \cite{Jorgensen-etal1996C} when modeling the number  of hospital emergency visits and in \cite{Botter-etal-2002} when modeling the mortality by lung diseases a latent (unobservable) stochastic process represented a time varying morbidity. These stochastic process based constructions are possible if the distributions of the increments of processes are infinite divisible, see \cite{Jorgensen-etal1996A} for details, which is the case for the EDMs that have the index set $\Lambda$  equal to $\re_+$. The infinite divisibility condition is satisfied by the Tweedie exponential dispersion models.

% ============================================================= %
\subsubsection{Estimating Equations Inference}
% ============================================================= %

\cite{Zeger-Liang} considered the generalized estimating equations (GEEs) to analyze longitudinal data based on quasi-likelihood methods. \cite{Liang-Zeger} derived the GEEs from a different and slightly more limited context. The method derives from EDMs, but is essentially
based on second-moment assumptions for the response. In both articles, the GEEs are derived without fully specifying the joint distribution. The regression coefficients are consistently estimated even when the correlation structure is misspecified. However, efficiency depends on the working correlation matrices proximity to the true one \citep{Liang-Zeger}. Liang and Zeger's method has been widely used in several areas dealing with non-Gaussian correlated data \citep{Hardin-Hilbe}.

\cite{Artes-Jorgensen} extended the GEE method to the class of dis\-persion models to handle certain types of non-normal data such as angles and proportions that are not well accommodated by EDMs, and for which there are currently no good methods avai\-la\-ble for longitudinal data analysis. 

\cite{Song-Tang} proposed methods to directly model the marginal means of the longitudinal proportional responses using the simplex distribution that takes into account the fact that such responses are percentages restricted between zero and one and may as well have large dispersion.

% ============================================================= %
\section{Concluding Remarks and Future Perspectives}
% ============================================================= %

As mentioned above, the Danish statistician Bent J\o{}rgensen (April 15, 1954; November 19, 2015) made several vital contributions in the area of statistical modeling.  He supervised many students in Denmark, Brazil,  and Canada and developed a vast international scientific collaboration network. Bent worked on a combination of theoretical and applied topics, including exponential families, univariate and multivariate dispersion models, exponential dispersion models, proper dispersion models, Tweedie distributions, generalized estimating equations and other types of statistical models. Although Bent's domain was mainly theoretical statistics, he also made significant contributions in a wide range of applied fields such as insurance, meteorology, and marine ecology, among others. In this article, as colleagues and friends, we outlined some details of his carrier and reviewed some of his main contributions, especially, univariate dispersion models, exponential dispersion and proper dispersion models.

The work of Bent J\o{}rgensen opened new research areas and inspired other researchers in the field. As a piece of evidence, we mention higher-order asymptotics  (see Section \ref{SS.IT}), the field of exponential family non-linear regressions (see Section \ref{SS.EFNLR}) and the latent stochastic processes based models (see Section \ref{SS.latentProcesses}). We envision that further developments might appear by expanding the general theory of dispersion models to different multivariate and dependent observation contexts (already partially done).  Defining general dispersion models via integral transforms other than the Laplace transform might allow incorporating heavy tail distributions and different types of stochastic processes not considered yet.

% ============================================================= %
\section*{Acknowledgements}
% ============================================================= %

We thank Jeanett S. Pelck (Applied Statistics Laboratory, Department of Mathematics, Aarhus University) and Ole Barndorff-Nielsen (Department of Mathematics, Aarhus University) for helpful comments which improve the manuscript. We are also grateful to the National Council for Scientific and Technological Development (CNPq) and the National Council for the Improvement of Higher Education (CAPES) for financial support of the first and the third authors.

% ============================================================= %

\end{document}